\documentclass[epjCONF]{svjour}

\usepackage{graphicx}
\usepackage[varg]{txfonts}
\usepackage[latin1]{inputenc}

\session-title{%
19$^{\textnormal{\footnotesize th}}$ International %
IUPAP Conference on Few-Body Problems in Physics%
}

\newcommand{\Order}[1]{\mathcal{O}#1}
\newcommand{\Lagr}[1]{\mathcal{L}#1}
\newcommand{\eqref}[1]{(\ref{#1})}

\newcommand{\beq}{\begin{equation}}
\newcommand{\eeq}{\end{equation}}
\newcommand{\bea}{\begin{eqnarray}}
\newcommand{\eea}{\end{eqnarray}}
\newcommand{\Mpi}{M_\pi^2}
\newcommand{\Mc}{M_{\pi^+}}
\newcommand{\Mn}{M_{\pi^0}}
\newcommand{\M}{M_{\pi}}
\newcommand{\p}{{\bf p}}
\newcommand{\eps}{\epsilon}

\begin{document}

\title{Cusp effects in meson decays}
\author{Bastian Kubis\inst{}\thanks{\email{kubis@hiskp.uni-bonn.de}}}

\institute{Helmholtz-Institut f\"ur Strahlen- und Kernphysik (Theorie) and
Bethe Center for Theoretical Physics, Universit\"at Bonn, D-53115 Bonn, Germany
}

\abstract{
The pion mass difference generates a pronounced cusp in the $\pi^0\pi^0$
invariant mass distribution of  $K^+\to\pi^0\pi^0\pi^+$ decays.
As originally pointed out by Cabibbo, an accurate measurement of the cusp
may allow one to pin down the S-wave pion--pion scattering lengths to high
precision.  We present the non-relativistic effective field theory
framework that permits to determine the structure of this cusp in a
straightforward manner, including the effects of radiative corrections.
Applications of the same formalism to other decay
channels, in particular $\eta$ and $\eta'$ decays, are also discussed.
} 

\maketitle

\section{The pion mass and pion--pion scattering}
\label{intro}

\begin{sloppypar}
The approximate chiral symmetry of the strong interactions 
severely constrains the properties and interactions 
of the lightest hadronic degrees of freedom, the
would-be Goldstone bosons (in the chiral limit of vanishing 
quark masses) of spontaneous chiral symmetry breaking that can be identified with the pions.  
The effective field theory that systematically exploits all the consequences
that can be derived from symmetries is chiral perturbation theory~\cite{weinbergchpt,glannphys}, 
which provides an expansion of low-energy observables in terms of small quark masses
and small momenta.  

One of the most elementary consequences of chiral symmetry is the well-known 
Gell-Mann--Oakes--Renner relation~\cite{GMOR} for the pion mass $M$ in terms of the light quark masses
(at leading order),
\beq
M^2 = B(m_u+m_d) ~,\quad  B = - \frac{\langle 0| \bar uu| 0 \rangle}{F^2} ~.
\eeq
A non-vanishing order parameter $B$, related to the light quark condensate via the 
pion decay constant $F$ (in the chiral limit), is a sufficient (but not necessary)
condition for chiral symmetry breaking.  Chiral perturbation theory allows to calculate
corrections to this relation~\cite{glannphys},
\beq
\Mpi = M^2 - \frac{M^4}{32\pi^2 F^2} \bar\ell_3 + \Order(M^6) ~, \label{eq:l3}
\eeq
with the a priori unknown low-energy constant $\bar\ell_3$.  Another way to write
Eq.~\eqref{eq:l3} is therefore
\beq
\Mpi = B(m_u+m_d) + A(m_u+m_d)^2 + \Order(m_{u,d}^3) ~,
\eeq
and the natural question arises: how do we know that the leading term in the 
quark-mass expansion of $\Mpi$ really dominates the series?  $\bar\ell_3$ could actually
be anomalously large, the consequence of which has been explored as an alternative scenario of chiral
symmetry breaking under the name of \emph{generalized} chiral perturbation
theory~\cite{genChPT}.

Fortunately, chiral low-energy constants tend to appear in more than one observable, 
and indeed, $\bar\ell_3$ also features in the next-to-leading-order corrections to the
isospin $I=0$ S-wave pion--pion scattering length 
$a_0^0$~\cite{glannphys},
\bea
a_0^0 &=& \frac{7\Mpi}{32\pi F_\pi^2} \big\{ 1+ \eps + \Order(M_\pi^4) \big\} ~, \nonumber\\
\eps &=&  \frac{5\Mpi}{84\pi^2 F_\pi^2} \Big( \bar\ell_1 + 2 \bar\ell_2 
- \frac{3}{8} \bar\ell_3 + \frac{21}{10} \bar\ell_4  + \frac{21}{8} \Big) ~.\label{eq:a00}
\eea
$\bar\ell_1$ and $\bar\ell_2$ are known from $\pi\pi$ D-waves $a_2^0$, $a_2^2$, while $\bar\ell_4$
can be determined from a dispersive analysis of the scalar radius of the pion 
$\langle r^2\rangle_\pi^S$~\cite{Donoghue,CGL2}, such that the correction term $\eps$
in Eq.~\eqref{eq:a00} can be rewritten as
\beq
\eps = \Mpi \left\{ \frac{\langle r^2\rangle_\pi^S}{3}
+ \frac{200\pi}{7}F_\pi^2 \left(a_2^0 + 2a_2^2\right) 
 - \frac{15\bar\ell_3-353}{672\pi^2 F_\pi^2} \right\} ~.\label{eq:epsa00}
\eeq
Consequently, a measurement of $a_0^0$
can lead to a determination of $\bar\ell_3$, and hence to a clarification of the
role of the various order parameters of chiral symmetry breaking in nature.
We wish to point out that Eq.~\eqref{eq:epsa00} only rewrites the dependence of $a_0^0$ 
on the $\Order(p^4)$ low-energy constants $\bar\ell_{1-4}$ in the form of a low-energy theorem.
The theoretical predictions of the two S-wave $\pi\pi$ scattering lengths
of isospin 0 and 2
from a combination of two-loop chiral perturbation theory~\cite{CGL1,CGL2}
and a Roy equation analysis~\cite{ACGL} (for QCD in the isospin limit),
\bea
a_0^0 &=& 0.220 \pm 0.005 ~, \nonumber\\
a_0^2 &=& -0.0444 \pm 0.0010 ~,\nonumber\\
a_0^0-a_0^2 &=& 0.265 \pm 0.004 ~,\label{eq:pipiRoy}
\eea
do not depend on the D-wave scattering lengths as input, but rather yield values for
all $\pi\pi$ threshold parameters as results.
The predictions Eq.~\eqref{eq:pipiRoy} are among the most precise in low-energy hadron
physics and present a formidable challenge for experimental verification.
For other recent phenomenological determinations of the scattering lengths, 
see Refs.~\cite{descotes,yndurain,kaminski}.

Traditionally, information on pion--pion scattering has been extracted from reactions 
on nucleons, which is difficult to achieve in a model-independent way, and 
data are usually 
not available very close to threshold kinematics.  
The latest precision determinations therefore mainly concern three different methods:
the lifetime measurement of pionium~\cite{DIRAC},
$K_{e4}$ decays~\cite{Pislak:Ke4,Batley:Ke4},
and, most recently, the so-called cusp effect in $K\to 3\pi$ decays.

Let us very briefly discuss the first two modern experimental approaches.
Pionium is the electromagnetically bound state of a $\pi^+\pi^-$ pair, with an ionization energy
of about 1.86~keV and a ground state width of about 0.2~eV.
Its energy levels as given by purely electromagnetic binding are perturbed by the 
short-ranged strong interactions:
they are shifted by elastic strong rescattering $\pi^+\pi^-$, 
but in particular, even the ground state is not stable, it decays dominantly into $\pi^0\pi^0$.
The decay width is given by the following (improved) Deser formula~\cite{Deser,Gall99}
\beq
\Gamma = \frac{2}{9} \alpha^3 p \, \big| a_0^0 - a_0^2\big|^2 (1+\delta) ~, 
\eeq
where $\alpha$ is the fine structure constant,
$p$ the momentum of a final-state $\pi^0$ in the center-of-mass frame,
and $\delta$ is a numerical correction factor accounting for isospin violation
beyond leading order, $\delta = 0.058 \pm 0.012$~\cite{GLRG}.  
Given the theoretical values for the $\pi\pi$ scattering lengths
of Eq.~\eqref{eq:pipiRoy}, the pionium lifetime can be predicted to be
\beq 
\tau = (2.9 \pm 0.1) \times 10^{-15} s ~, \label{eq:pioniumlife}
\eeq
while ultimately, the argument should be reversed, and a measurement of the lifetime
is to be used for a determination of $a_0^0-a_0^2$.  
The current value from the DIRAC experiment~\cite{DIRAC},
\beq \tau = \left(2.91^{+0.49}_{-0.62}\right)\times 10^{-15} s ~,
\eeq
agrees with Eq.~\eqref{eq:pioniumlife}, but is not yet comparably precise.
For a comprehensive review of the theory of hadronic atoms, see Ref.~\cite{HadAtom}.
\end{sloppypar}

The decay $K^+\to\pi^+\pi^-e^+\nu_e$ ($K_{e4}$) can be described in terms of hadronic
form factors, which, in the isospin limit, share the phases of $\pi\pi$ scattering 
due to Watson's final state theorem~\cite{Watson}.  
What can be extracted unambiguously from the decay,
using the so-called Pais--Treiman method~\cite{PaisTreiman},
is the difference of $\pi\pi$ $I=0$ S-wave and $I=1$ P-wave phase shifts
\beq
\delta_0^0(s_{\pi\pi}) - \delta_1^1(s_{\pi\pi}) ~,
\eeq
and as the energy of the two pions is kinematically restricted to $\sqrt{s_{\pi\pi}}<M_K$,
these phases are accessible close to threshold.
It has been pointed out~\cite{CGR:Ke4} that, given the precision of the latest
NA48/2 data~\cite{Batley:Ke4}, it is necessary to include an isospin-breaking correction
phase in the analysis.  The resulting scattering length determination will be shown
in the comparison in Sect.~\ref{sec:radcorr}.

\section{The cusp effect in \boldmath{$K^\pm\to\pi^0\pi^0\pi^\pm$} decays}\label{sec:cusp}

\begin{figure}
\includegraphics[width=\linewidth]{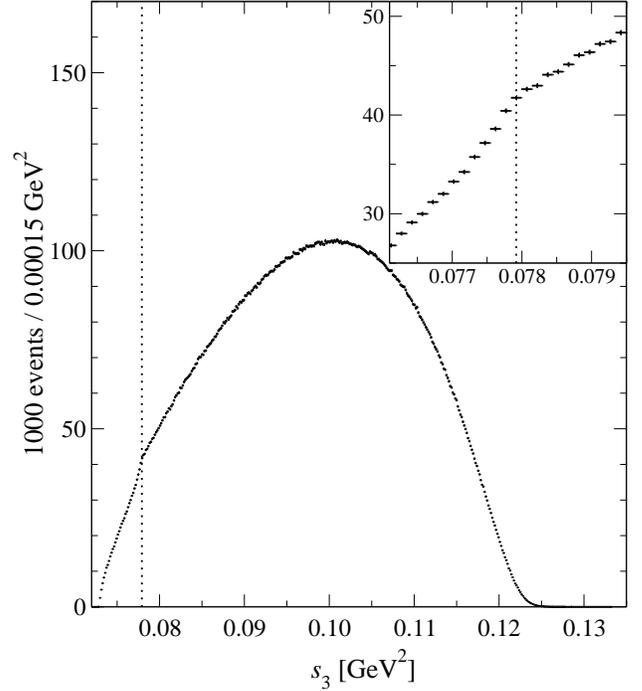}
\caption{Cusp in the decay spectrum $d\Gamma/ds_3$ of the decay $K^\pm\to\pi^0\pi^0\pi^\pm$
as seen by the NA48/2 collaboration.  The dotted vertical line marks the position of 
the $\pi^+\pi^-$ threshold, the insert focuses on the cusp region.
Data taken from Ref.~\cite{NA48}.}
\label{fig:expcusp}
\end{figure}
\begin{figure}
\includegraphics[width=\linewidth]{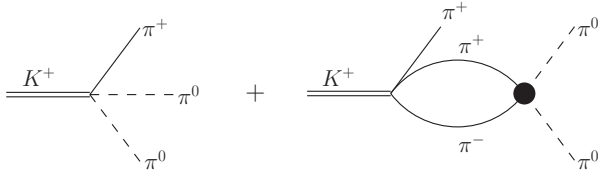}
\caption{``Direct'' and ``rescattering'' contribution to the decay $K^+\to\pi^0\pi^0\pi^+$.
The black dot marks the charge-exchange $\pi\pi$ scattering vertex proportional to 
the scattering lengths at threshold.}
\label{fig:direct+cex}
\end{figure}

In an investigation of the decay $K^\pm\to\pi^0\pi^0\pi^\pm$, the NA48/2 collaboration at CERN
has observed a cusp, i.e.\ a sudden, discontinuous change in slope, in the decay spectrum
with respect to the invariant mass squared of the $\pi^0\pi^0$ pair
$d\Gamma/ds_3$, $s_3=M_{\pi^0\pi^0}^2$~\cite{NA48}; see Fig.~\ref{fig:expcusp}.  
A first qualitative explanation was subsequently given by Cabibbo~\cite{Cabibbo}, who 
pointed out that a $K^+$ can, simplistically speaking, either decay ``directly'' into the 
$\pi^0\pi^0\pi^+$ final state, or alternatively decay into three charged pions $\pi^+\pi^+\pi^-$, 
with a $\pi^+\pi^-$ pair rescattering via the charge-exchange process into two neutral pions, 
compare Fig.~\ref{fig:direct+cex}.
The loop (rescattering) diagram has a non-analytic piece proportional to
\beq
i\,v_\pm(s_3) =
 \left\{
\begin{array}{ll}
i \sqrt{1-\frac{4\Mc^2}{s_3}} ~, & s_3>4\Mc^2 ~, \\
- \sqrt{\frac{4\Mc^2}{s_3}-1} ~, & s_3<4\Mc^2 ~,
\end{array} 
\right. 
\eeq
and as the charged pion is heavier than the neutral one by nearly 4.6~MeV, 
the (then real) loop diagram can interfere with the ``direct'' decay 
below the $\pi^+\pi^-$ threshold and produce a square-root-like singularity 
at $s_3=4\Mc^2$, the cusp visible in the experimentally measured spectrum Fig.~\ref{fig:expcusp}. 
Such threshold singularities have of course been known for a long time~\cite{Wigner}
and have been re-discovered for the scattering of neutral pions in the context
of chiral perturbation theory~\cite{MMS}.  
It had even been pointed out very early by Budini and Fonda~\cite{fonda} that these cusps may be used to investigate
$\pi\pi$ scattering: as suggested in Fig.~\ref{fig:direct+cex}, 
the strength of the cusp is proportional to the charge-exchange pion--pion scattering amplitude
at threshold, hence a combination of scattering lengths might be extracted from a precision
analysis of the cusp effect.  

The challenge for theory is to provide a framework that matches the tremendous accuracy 
of the experimental data: the partial data sample analyzed in Ref.~\cite{NA48} was based
on $2.3\times 10^7$ $K^\pm\to\pi^0\pi^0\pi^\pm$ decays, subsequently expanded to 
more than $6.0\times 10^7$ decays~\cite{CuspEPJC}.
Different theoretical approaches have been suggested to this end:
a combination of analyticity and unitarity with an expansion of the rescattering effects
in powers of the $\pi\pi$ threshold parameters~\cite{CI}, and chiral perturbation theory beyond one-loop
order~\cite{GPS}.  In the following, we advocate the use of non-relativistic effective
field theory~\cite{CGKR} as the appropriate systematic tool to analyze these decays.

\section{Non-relativistic effective field theory}

\begin{sloppypar}
Consider as a starting point a generic $\pi\pi$ partial wave amplitude $T$.  
Close to threshold, its real part can be written in the effective range expansion according to
\beq
{\rm Re}\, T = a + b \,q^2 + c \,q^4 + \ldots ~,
\eeq
with the scattering length $a$, the effective range $b$, a shape parameter $c$ etc.
Chiral perturbation theory allows to calculate the parameters $a$, $b$, $c$ 
to a certain accuracy in the quark-mass (or pion-mass) expansion, see Eq.~\eqref{eq:a00} 
for an example, but in principle, each of these parameters receives contributions
from each loop order.
On the other hand, one can set up a non-relativistic effective field theory (NREFT) in such a way that the scattering
length $a$ is entirely given in terms of tree graphs, without any further loop corrections;
similarly, the effective range $b$ can be calculated from tree and two-loop graphs only, 
but then no further contributions.  In other words, NREFT allows to parameterize $T$ \emph{directly}
in terms of threshold parameters.  Note that this is exactly what we want: the aim here is not
to predict the scattering lengths, but to provide a representation of the (scattering or decay) 
amplitude in terms of the $\pi\pi$ threshold parameters that allows for an accurate extraction
of the latter from experimental data.
This is similar in spirit to the use of NREFT in the analysis of hadronic atoms in order
to extract scattering lengths of different systems from their life times or energy level
shifts (see Ref.~\cite{HadAtom} and references therein).

\subsection{Power counting, Lagrangians}\label{sec:power}

First, we need to specify our power counting scheme.
We introduce a formal non-relativistic parameter $\eps$ and count 3-momenta of the pions
in the final state according to $|\vec{p}|/\M = \Order(\eps)$.  Consequently, 
the pions' kinetic energies are 
\beq 
T_i = \omega_i(\p_i)-M_i = \Order(\eps^2) ~,~ 
\textrm{where}~~ \omega_i(\p_i) = \sqrt{M_i^2+\p_i^2}~,
\eeq
with $i=1,\,2,\,3$, $M_1=M_2=\Mn$, $M_3=\Mc$, and the $Q$-value of the reaction has to 
be counted as $\Order(\eps^2)$, too, as 
\beq
M_K - \sum_i M_i = \sum_i T_i = \Order(\eps^2) ~.
\eeq
In addition, we adopt the suggestion of Ref.~\cite{CI}: as the $\pi\pi$ scattering lengths
are small due to the Goldstone nature of the pions, their final-state rescattering can 
be taken into account perturbatively, in contrast to what one has to do e.g.\ 
in the treatment of three-nucleon systems.
Hence in this case, we can make use of a two-fold expansion in $\eps$ 
and $a$, by which we generically denote all $\pi\pi$ threshold parameters. 
This scheme allows for a consistent power counting in the sense that at any given order
in $a$ and $\eps$, only a finite number of graphs contributes.

The polynomial terms contributing at tree level are organized in even powers of momenta, 
hence there are terms of order $\eps^0$, $\eps^2$, $\eps^4$, \ldots.
\begin{figure}
\includegraphics[width=\linewidth]{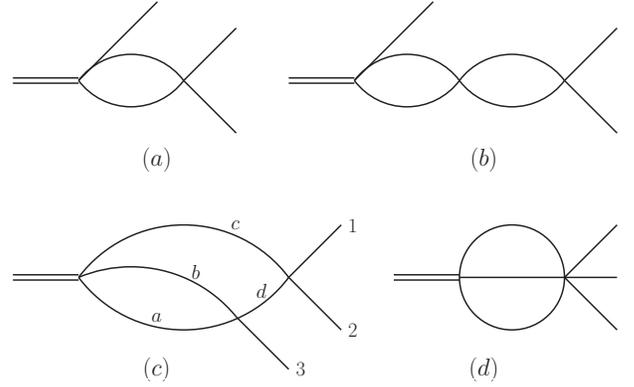}
\caption{Topologies for $\pi\pi$ rescattering graphs at one and two loops.
The double line denotes the decaying kaon, while single lines stand for either charged or neutral pions.}
\label{fig:topologies}
\end{figure}
Slightly more complicated is the power counting of the loop graphs, of which 
the typical topologies at one- and two-loop order are shown in Fig.~\ref{fig:topologies}.
Generically, the non-relativistic pion propagators are of a form
\beq
\propto \frac{1}{\omega(\p)-p^0} = \Order(\eps^{-2})
\eeq 
(note however the discussion below on its precise form), 
while a loop integration is counted according to 
\beq 
d^4 p = dp^0 d^3\p = \Order(\eps^5) ~.
\eeq
Consequently, we find that each additional loop induced by two-body rescattering is
suppressed by a factor of 
\beq 
(\eps^{-2})^2 \,\eps^5 = \eps ~,
\eeq
such that the one-loop diagram (a) in Fig.~\ref{fig:topologies} is of $\Order(a^1\eps^1)$, 
while the two two-loop graphs (b) and (c) are of $\Order(a^2\eps^2)$.  We therefore find a \emph{correlated}
expansion in $a$ and $\eps$: loops are not only suppressed by powers of the $\pi\pi$ 
threshold parameters, but in addition by powers of $\eps$.

Finally, a rescattering graph due to three-body interactions that first appears at
two-loop order, see diagram (d) in Fig.~\ref{fig:topologies}, is suppressed by
\beq
(\eps^{-2})^3 (\eps^5)^2 = \eps^4 ~.
\eeq  
This diagram is not proportional to any $\pi\pi$ 
threshold parameters; however, the graph is a constant, and its main effect apart 
from coupling constant renormalization is to give the $K\to3\pi$ vertex a small 
imaginary part.  Specifically, denoting the leading $\Order(\eps^0)$ $K^+\to\pi^0\pi^0\pi^+$
vertex by $G_0$ (see Eq.~\eqref{eq:Ls} below), the $\pi^0\pi^0\pi^+$ intermediate state,
with elastic three-particle rescattering approximated by the leading-order vertex derived from 
chiral perturbation theory, leads to an imaginary part of
\bea
\frac{{\rm Im}\, G_0}{{\rm Re}\, G_0} &=& 
\frac{(M_K-3\M)^2}{256\pi^2}\,\frac{M_\pi^2}{24\sqrt{3}F_\pi^4} + \Order\big((M_K-3\M)^3\big) \nonumber\\
&\simeq& 1.5\cdot 10^{-5} ~,
\eea
which therefore indeed turns out to be formally of $\Order(\eps^4)$, but numerically 
entirely negligible.

As the cusp effect depends essentially on the analytic properties of the amplitude,
it is clearly desirable to preserve the latter exactly, i.e.\ to correctly reproduce the
singularity structure of the relativistic decay amplitude in the low-energy region
$|\p| \ll \M$; only far-away singularities associated e.g.\ with the creation and annihilation
of particle--antiparticle pairs (inelastic channels) should be subsumed in effective
coupling constants.  To this end, we use a pion propagator of the form
\beq
\frac{1}{2\omega(\p)}\frac{1}{\omega(\p)-p^0} ~.\label{eq:NRprop}
\eeq
This corresponds to the complete particle-pole piece of the full relativistic propagator,
\beq
\frac{1}{\M^2-p^2} = \frac{1}{2\omega(\p)}\frac{1}{\omega(\p)-p^0} + \frac{1}{2\omega(\p)}\frac{1}{\omega(\p)+p^0} ~,
\eeq
and therefore reproduces the correct relativistic dispersion law.  The propagator Eq.~\eqref{eq:NRprop}
can be generated by a non-local kinetic-energy Lagrangian
\beq
\Lagr_{\rm kin} = \Phi^\dagger (2W)(i\partial_t-W)\Phi ~, \quad W=\sqrt{\M^2-\Delta} ~,
\eeq
where $\Phi$ represents the pion field operator and $\Delta$ is the Laplacian.
$\Lagr_{\rm kin}$ generates all relativistic corrections in the propagator and leads to a 
manifestly Lorentz-invariant and frame-independent amplitude.

In order to restore the naive power counting rules for loop graphs in the presence
of explicit heavy (pion) mass scales, one has to apply the
threshold expansion~\cite{Beneke,GLRG,HadAtom}: all loop integrals are expanded
in powers of the inverse pion mass, integrated order by order, and the results subsequently resummed.
In particular the presence of the square roots $\omega(\p)$ in the propagator Eq.~\eqref{eq:NRprop}
leads to significant technical complications in the calculation of the loops.

We need two types of interaction terms in the effective Lagrangian, generating the $\pi\pi$ interaction
as well as the $K\to3\pi$ tree level amplitudes,
\bea
\Lagr_{\pi\pi} &=& C_x \bigl( \Phi_-^\dagger \Phi_+^\dagger (\Phi_0)^2 + h.c. \bigr) + \ldots +\Order(\eps^2) ~, \label{eq:Ls} \\
\Lagr_{K3\pi} &=& \frac{G_0}{2} K_+^\dagger \Phi_+ (\Phi_0)^2 +
                \frac{H_0}{2} K_+^\dagger \Phi_- (\Phi_+)^2 + h.c. + \Order(\eps^2) ~,\nonumber
\eea
where we have only displayed the leading, energy-in\-dependent couplings, and the ellipsis in $\Lagr_{\pi\pi}$
denotes similar interaction terms for the other possible $\pi\pi$ scattering channels.
The current accuracy of the calculation of the $K\to3\pi$ decay amplitude includes all terms
up to and including $\Order(a^0\eps^4,a^1\eps^5,a^2\eps^4)$; for this purpose, 
$\Lagr_{K3\pi}$ as well as S-wave interaction terms in $\Lagr_{\pi\pi}$
are needed up to $\Order(\eps^4)$, 
while only P-wave scattering lengths and no D-wave contributions are necessary in $\Lagr_{\pi\pi}$.
\end{sloppypar}

The whole framework briefly sketched here is a completely Lagrangian-based quantum field theory, 
hence all constraints from analyticity and unitarity are automatically obeyed.

\subsection{Matching}

\begin{sloppypar}
The coupling constants of the non-relativistic Lagrangians Eq.~\eqref{eq:Ls} have to be related to physical observables
in the underlying relativistic field theory.
In the case of the couplings of $\Lagr_{\pi\pi}$, they can be matched using the effective range expansion
of the $\pi\pi$ scattering amplitude.  For example, $C_x$ as defined in Eq.~\eqref{eq:Ls} is
related to the charge-exchange amplitude $T_x = T(\pi^+\pi^-\to\pi^0\pi^0)$ by
\bea
{\rm Re}\, T_x &=& 2C_x + \Order(\eps^2) ~, \nonumber \\
2C_x &=& -\frac{32\pi}{3}\left(a_0^0-a_0^2\right)\biggl\{ 1+ \frac{\Mc^2-\Mn^2}{3\Mpi} \biggr\} + \Order(e^2p^2) \nonumber\\
&=& -\frac{32\pi}{3}\left(a_0^0-a_0^2\right)\bigl\{ 1+ (0.61\pm 0.16)\times10^{-2} \bigr\} \nonumber\\ && + \Order(e^2p^4) ~. 
\label{eq:pipiMatch}
\eea
The isospin-breaking corrections in relating the charge-exchange amplitude at threshold to the scattering lengths
of definite isospin are calculated in chiral perturbation theory, where the second line in Eq.~\eqref{eq:pipiMatch} shows the analytic
correction at $\Order(e^2)$, while the numerical estimate in the third line includes the higher order of
$\Order(e^2p^2)$~\cite{KnechtUrech,GLRG}.
\end{sloppypar}

The polynomial terms $G_0$, $G_1$, $\ldots$ for $K^+ \to \pi^0\pi^0\pi^+$ and $H_0$, $H_1$, $\ldots$ 
for $K^+ \to \pi^+\pi^+\pi^-$ are not strictly matched, but used as a parameterization of the amplitudes in question.
They replace the more traditional Dalitz plot parameters used for that purpose in experimental fits
neglecting non-trivial final-state rescattering effects.
The strategy is to fit (in principle) all parameters of the non-relativistic representation to data
of both decay channels, and then determine the scattering length combination $a_0^0-a_0^2$
via Eq.~\eqref{eq:pipiMatch}.
In practice, one may decide to use some of the parameters, for instance higher-order $\pi\pi$ threshold parameters
such as effective ranges or P-waves, as input, employing their theoretically predicted values~\cite{CGL2} instead.

\subsection{Analytic structure of the non-trivial \\\hspace{5.9mm}two-loop graph}

The function $F(s)$ describing the non-trivial, genuine two-loop graph
(c) in Fig.~\ref{fig:topologies}
can be expressed analytically in terms of logarithms
(see Ref.~\cite{KL} for an explicit closed representation).  
Close to threshold, it can be approximated according to
\beq
 F(s) \simeq  
\frac{v_\pm(s)}{256\pi^2}\sqrt{\frac{M_K^2-9\M^2}{M_K^2-\M^2}} \label{eq:Fthresh}
\eeq
(for all pion masses running in the loop equal),
which is manifestly of $\Order(\eps^2)$ as required by the power counting
set up in Sect.~\ref{sec:power}.
However, a decomposition of the full two-loop function according to
\beq
F(s) = A(s) + B(s) \, v_\pm(s) ~, \label{eq:decompAB}
\eeq
with both $A(s)$, $B(s)$ analytic functions of $s$ as suggested in Ref.~\cite{CI}, 
turns out not to hold.  In fact, it can be shown~\cite{forthcoming} that if one enforces such a decomposition,
both $A(s)$ and $B(s)$ diverge at the border of phase space for maximal $s$ like $1/\sqrt{s_p-s}\,$, $s_p=(M_K-M_3)^2$, 
in such a way that the sum $A(s) + B(s) \, v_\pm(s)$ is finite.

What is more, at least for certain pion mass assignments within the loop,
the decomposition Eq.~\eqref{eq:decompAB} even fails as a representation
of the analytic structure of $F(s)$ within the decay region.
With the pions in the loop labelled as indicated in Fig.~\ref{fig:topologies}, 
the solutions of the Landau equations~\cite{landau,smatrix} show that
anomalous thresholds exist for
\bea
s^\pm &=& \frac{1}{2} \biggl\{ M_K^2 +M_3^2+M_c^2+M_d^2-(M_a+M_b)^2 \nonumber\\
&&\qquad + \frac{(M_K^2-M_c^2)(M_d^2-M_3^2)\pm \sqrt{\lambda_1\lambda_2}}{(M_a+M_b)^2} \biggr\} ~, \nonumber\\
\lambda_1&=&\lambda\left((M_a+M_b)^2,M_3^2,M_d^2\right) ~, \nonumber\\
\lambda_2&=&\lambda\left(M_K^2,(M_a+M_b)^2,M_c^2\right) ~, \label{eq:anomthresh}
\eea
with the standard K\"all\'en function $\lambda(a,b,c)=a^2+b^2+c^2-2(ab+ac+bc)$.
\begin{figure}
\includegraphics[width=\linewidth]{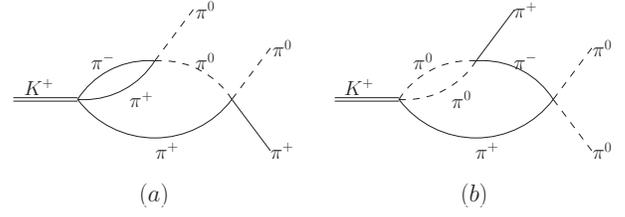}
\caption{Two-loop diagrams for $K^+\to\pi^0\pi^0\pi^+$ displaying anomalous thresholds.
Double, single, and dashed lines stand for charged kaons, charged, and neutral pions, respectively.}
\label{fig:anomthresh}
\end{figure}
According to Eq.~\eqref{eq:anomthresh}, the analytic structure of $F(s)$ is particularly
intricate for $M_a+M_b \neq M_3+M_d$.  The two relevant graphs in $K^+\to\pi^0\pi^0\pi^+$ fulfilling this condition
are shown in Fig.~\ref{fig:anomthresh}.  
In the case of diagram (a), the $s_1^\pm$ are \emph{real} and yield branch points in the amplitude
at $\sqrt{s_1^-}=308\,$MeV and $\sqrt{s_1^+}=356\,$MeV, compared to the phase space limits in this variable, 
given by threshold $\sqrt{s_t}=275\,$MeV and pseudothreshold $\sqrt{s_p}=359\,$MeV.
The functions $A(s)$ and $B(s)$ in the decomposition Eq.~\eqref{eq:decompAB}
display singular behavior at $s_1^\pm$, 
while the complete amplitude is analytic at the upper rim of the cut
(that starts at $s_t=(\Mc+\Mn)^2$); it becomes singular only at its lower rim.  
Diagram (b), on the other hand, has \emph{complex} anomalous thresholds $s_3^\pm$,
which lead to special complications in a dispersive representation of this loop graph,
necessitating a deformation of the integration path over the discontinuity.  
Again, on the upper rim of the cut, the complete loop function $F(s)$ is analytic
also in that case.

We wish to emphasize once more that the singularity structure of the explicit non-relativistic
representation of $F(s)$~\cite{KL} in the low-energy region, i.e.\ in the decay region and slightly
beyond, is identical to the one of the fully relativistic amplitude. 
It therefore includes all the anomalous thresholds discussed above, 
in precisely the right kinematical positions.

\subsection{Two-loop representation}

\begin{sloppypar}
The full representation of the $K\to3\pi$ decay amplitudes up to and including $\Order(a^0\eps^4,a^1\eps^5,a^2\eps^4)$
comprises tree, one-loop, and two-loop graphs of the topologies shown in Fig.~\ref{fig:topologies}, 
with all possible charge combinations of intermediate pions.
The only loop function at one loop, see graph (a) in Fig.~\ref{fig:topologies}, is given by
\beq
J(s) = \frac{i\,v(s)}{16\pi} ~, \label{eq:J}
\eeq
hence produces precisely the analytic structure discussed in Sect.~\ref{sec:cusp}.
The two-loop graph (b) in Fig.~\ref{fig:topologies} is given as a product of 
two functions $J(s)$, hence it is real above threshold and, if it contains singular behavior
at $s_3=4\Mc^2$ (in a product of one ``charged'' and one ``neutral'' loop), the real square root
that interferes with the dominant tree graphs is also seen for $s_3>4\Mc^2$.  
The same is true for the threshold behavior of the genuine two-loop function discussed in the 
last section, see Eq.~\eqref{eq:Fthresh}.  Finally, the two-loop graph with three-body rescattering, 
diagram (d) in Fig.~\ref{fig:topologies}, is a constant and can essentially be absorbed 
in a redefinition of the tree-level couplings, hence it does not affect the analytic structure
in a non-trivial way.

The cusp up to two loops is therefore of the following generic structure:
while the one-loop diagrams generate a structure $\propto i\,a\,v_\pm(s_3)$ which interferes with
the (dominant) tree amplitude \emph{below} the $\pi^+\pi^-$ threshold (where the square root turns real), 
the two-loop graphs include terms $\propto a^2 v_\pm(s_3)$, hence a singular structure 
(in interference with the tree parts) \emph{above} that point. 
\begin{figure}
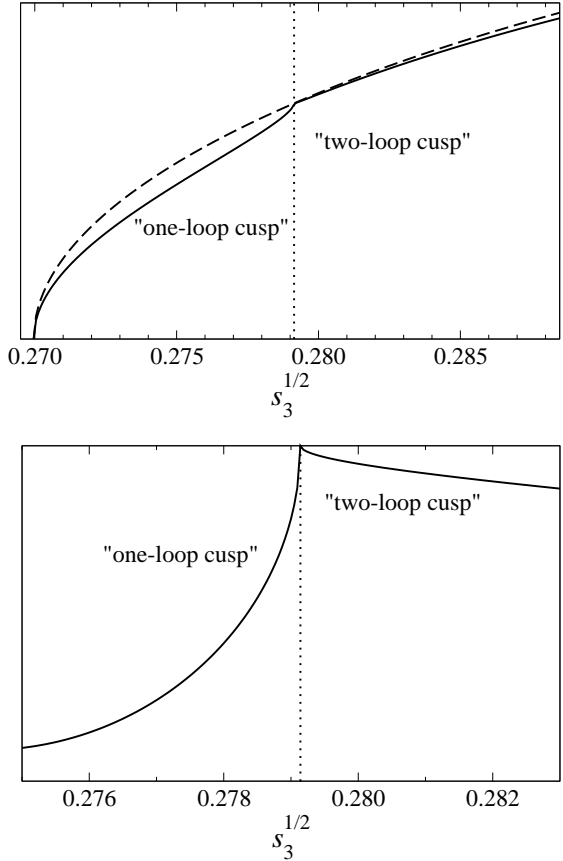
 \centering
\includegraphics[width=0.9\linewidth]{KubisB-one+twoloop_cusps.eps} \\[3mm] \hspace*{1mm}
\includegraphics[width=0.88\linewidth]{KubisB-cusp_diff.eps} 
\caption{Sketch of the cusps in the decay spectrum at $\Order(a)$ below, and at $\Order(a^2)$ above the 
$\pi^+\pi^-$ threshold, denoted by the vertical dotted line.
In the lower panel, focussing even closer on the threshold region,
the tree-level spectrum (dashed) is subtracted for better illustration of the small ``two-loop cusp''.}
\label{fig:one+twoloopcusps}
\end{figure}
This is illustrated schematically in Fig.~\ref{fig:one+twoloopcusps}.
The ``two-loop cusp'' above threshold is a much smaller effect, yet given the precision of the
data in the NA48/2 analysis, it has turned out to be a vital ingredient in the theoretical representation
in order to achieve a statistically adequate description.
As the cusp strength at two loops also incorporates $\pi\pi$ rescattering effects other than 
the charge exchange channel, there is in principle also (reduced) sensitivity to another linear
combination of S-wave scattering lengths, e.g.\ $a_0^2$ alone.  A fit of this form yields~\cite{CuspEPJC}
\bea
a_0^0-a_0^2 &=& 0.2815 \pm 0.0043_{\rm stat} \pm \ldots ~, \nonumber\\
a_0^2 &=& -0.0693 \pm 0.0136_{\rm stat} \pm \ldots ~. \label{eq:results_sans}
\eea
A comparison to the theoretical prediction Eq.~\eqref{eq:pipiRoy} shows that $a_0^0-a_0^2$ 
comes out uncomfortably large (by more than $3.5\sigma$). 
This turns out not to be a statistical accident, but there is a theoretical explanation for 
this discrepancy that we will discuss in the following section.

\section{Radiative corrections}\label{sec:radcorr}

Once the theoretical and experimental precision in the determination of hadronic, strong-interaction
physics observables arrives at the percent level, the effects of electromagnetic or radiative
corrections have to be taken into account.  
Such corrections in $K\to 3\pi$ decays have already been considered earlier in the framework of 
chiral perturbation theory~\cite{nehme,B1,B2},
or in a quantum-mechanical approach~\cite{tarasovk3piA,tarasovk3piB}. 
As we employ a Lagrangian framework, the inclusion 
of photons via minimal substitution is completely straightforward: 
\beq
\partial_\mu\Phi_\pm \to (\partial_\mu\mp ieA_\mu)\Phi_\pm  ~,
\eeq
and similarly for the kaons.  Furthermore, all possible 
non-minimal gauge invariant terms can be added.  
In the context of a non-relativistic theory, it is useful to work in the Coulomb gauge and 
differentiate between Coulomb and transverse photons, which feature differently in the generalized
power counting scheme.
In addition, for transverse photons, one has to differentiate between ``soft'' and ``ultrasoft'' modes:
while both have zero components that have to be counted according to $l_0=\Order(\eps^2)$, 
the three components are either $\vec{l}=\Order(\epsilon)$ for soft, or
$\vec{l}=\Order(\epsilon^2)$ for ultrasoft photons.
\begin{figure}
\includegraphics[width=\linewidth]{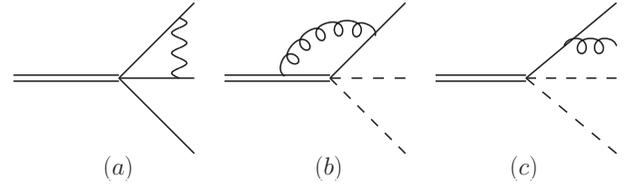}
\caption{Examples for diagrams with ``external'' radiative corrections.  
Double, single, and dashed lines stand for 
charged kaons, charged, and neutral pions, respectively.  
Wiggly and curly lines denote Coulomb and transverse photons, respectively.}
\label{fig:external}
\end{figure}
The summary of the counting rules for diagrams with virtual photons is then as follows~\cite{radcorr}:
\begin{enumerate}
\item Adding a Coulomb photon to a hadronic ``skeleton'' diagram modifies its counting by a factor of $e^2/\eps$.
An example of this is diagram (a) in Fig.~\ref{fig:external}: with a constant $K\to3\pi$ vertex of $\Order(\eps^0)$,
this diagram will scale as $\Order(e^2\eps^{-1})$, the negative power in $\eps$ indicating the presence
of the Coulomb pole in that graph.
\item Transverse photons couple to mesons with vertices of $\Order(\eps)$, hence soft transverse photons
are suppressed relative to Coulomb photon exchange by two orders in the $\eps$-expansion.
As an example, diagram (a) in Fig.~\ref{fig:external}, with the Coulomb photon replaced by a transverse one,
contributes at $\Order(e^2\eps)$.
\item Ultrasoft transverse photons added to a hadronic ``skeleton'' diagram finally 
change its power counting by a factor of $e^2\eps^2$.  As an example, the transverse photon in diagram (b)
of Fig.~\ref{fig:external} can be shown to be ultrasoft, hence with a constant $K\to3\pi$ vertex,
the graph scales as $\Order(e^2\eps^2)$.
\end{enumerate}

As is well known, the inclusion of \emph{virtual} photon effects requires the simultaneous
consideration of radiation of additional \emph{real} photons in order to obtain well-defined,
infrared-finite quantities. 
The observable that can be calculated including effects of $\Order(\alpha)$ 
(where $\alpha=e^2/4\pi$ is the fine structure constant) is~\cite{radcorr}
\bea
\frac{d\Gamma}{ds_3} \biggr|_{E_\gamma< E_{\rm max}} \hspace{-2mm} &=& 
\frac{d\Gamma(K\to 3\pi)}{ds_3} + \frac{d\Gamma(K\to 3\pi\gamma)}{ds_3} \biggr|_{E_\gamma<E_{\rm max}} \hspace{-2mm}
+\Order(\alpha^2) \nonumber\\
&=& \Omega(s_3, E_{\rm max}) \frac{d\Gamma^{\rm int}}{ds_3} + \Order(\alpha^2) ~.\label{eq:raddist}
\eea
The notation indicates that the emission of real or bremsstrahlung photons
is included up to a maximal energy $E_{\rm max}$, specified by the experimental detector resolution.
Here, the channel-dependent correction factor $\Omega(s_3, E_{\rm max})$ subsumes all ``external''
radiative corrections due to (real and virtual) corrections with the photons exclusively hooked
to charged external legs, see Fig.~\ref{fig:external} for examples,
\begin{figure}\centering
\includegraphics[width=\linewidth]{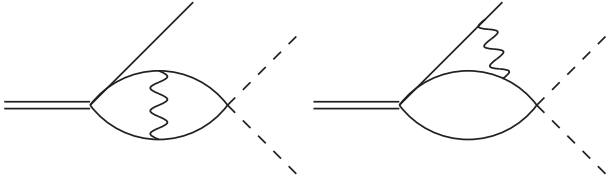}
\caption{Examples for diagrams with ``internal'' radiative corrections.  The wiggly lines denote
Coulomb photons; otherwise, see the line style in Fig.~\ref{fig:external}.}
\label{fig:internal}
\end{figure}
while $d\Gamma^{\rm int}/ds_3$ still includes other, ``internal'' radiative
corrections, see Fig.~\ref{fig:internal}.
In terms of the non-relativistic power counting, we aim for a calculation of the decay spectra
Eq.~\eqref{eq:raddist} including effects of $\Order(e^2\eps^4)$ for all $K\to3\pi$ channels. 
One has to bear in mind that 
the non-radiative decay spectrum $d\Gamma(K\to 3\pi)/ds_3$ starts at $\Order(\eps^2)$, while
the bremsstrahlung part $d\Gamma(K\to 3\pi\gamma)/ds_3$ begins to contribute at $\Order(e^2\eps^4)$, 
which means that it is sufficient to include radiation of real photons at leading order 
in the non-relativistic expansion.
Internal corrections of $\Order(e^2a^1\eps^2)$, on the other hand, are only included for 
the ``main'' decay channels displaying the cusp, $K^+\to\pi^0\pi^0\pi^+$ and $K_L\to3\pi^0$ 
(see Sect.~\ref{sec:KLeta} below).

The external radiative corrections are rather well-known (compare e.g.\ Ref.~\cite{isidorirad} 
for a relativistic approach); their effect on the decay spectrum Eq.~\eqref{eq:raddist} is small 
and smooth except for the Coulomb pole in channels with more than one charged particle in 
the final state.  These latter threshold singularities are usually taken care of by means of the 
Gamow--Sommerfeld factors~\cite{Gamow,Sommerfeld}
in the experimental analyses.
Even the soft-photon approximation for bremsstrahlung photons is rather 
accurate compared to the exact result~\cite{radcorr}.
In fact, the non-relativistic power counting and in particular the resulting powers in $\eps$
for the various radiative corrections nicely illustrate why the effects of Coulomb photons
in particular are important, while (finite) bremsstrahlung effects are very small.

\begin{figure}\centering
\includegraphics[width=0.55\linewidth]{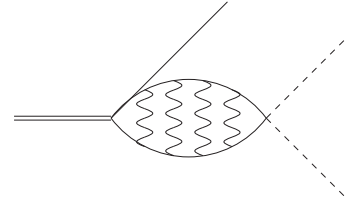}
\caption{Multi-photon exchange inside the charged-pion loop, responsible for pionium formation.}
\label{fig:pionium}
\end{figure}

In the context of the cusp analysis, the internal corrections are of potentially more intriguing
effect, as they modify the analytic structure of the decay amplitude near the $\pi^+\pi^-$ threshold.
These modifications become important as soon as $\alpha/v_\pm$ is not small any more.
The most remarkable effect is the formation of pionium due to multi-photon exchange inside
the charged-pionium loop, see Fig.~\ref{fig:pionium}, which leads to an infinite number of bound-state
poles close to threshold.
The analytic solution to the resummation of an infinite number of exchanged Coulomb photons is known
as the Schwinger Green's function~\cite{schwinger} replacing the simple one-loop function $J_\pm(s)$
in Eq.~\eqref{eq:J},
\bea
G(s) &=& \frac{i\,v_\pm(s)}{16\pi} \label{eq:pionium} \\
&-&\frac{\alpha}{16\pi} \biggl[ \frac{\log(-v_\pm^2(s))}{2}
+\Psi\Bigl(1-\frac{i\,\alpha}{2v_\pm(s)}\Bigr) - \Psi(1) + C \biggr] ~, \nonumber
\eea 
where $\Psi(x)=\frac{d}{dx}\log{\Gamma(x)}$, and $C$ is a constant.  The function $\Psi$ contains
the pionium pole terms, which are non-perturbative effects, with binding energies of $\Order(\alpha^2)$.
The leading effect however is, at $\Order(\alpha)$, the one-photon exchange, leading to the logarithmic
divergence at threshold $\propto \log(-v_\pm^2(s))$ in Eq.~\eqref{eq:pionium}.
Even if the central bin right at the cusp where pionium is formed is excluded in the experimental analysis,
such that the one-photon exchange is sufficient to describe the effects of electromagnetism, its
effects are surprisingly sizeable.  This is due to the fact that the cusp is precisely about a change
in the slope of the distribution: if its structure below threshold is, up to $\Order(\alpha)$, given by
\beq
-\frac{1}{16\pi^2} \Big\{ \sqrt{-v_\pm^2(s)} + \frac{\alpha}{2}\log\big(-v_\pm^2(s)\big)\Big\} ~,
\eeq
its derivative at a kinematical point $s=4\Mc^2-\Delta$ yields
\beq
\frac{1}{64\pi\M\sqrt{\Delta}} \biggl\{ 1 + \frac{2\alpha\M}{\sqrt{\Delta}} \biggr\} + \ldots~,
\eeq
where the ellipsis denotes higher orders in $\Delta$, such that the correction of $\Order(\alpha)$
becomes large near threshold.
Indeed, the fit to experimental data, including radiative corrections in the amplitude, yields
the following results for the $\pi\pi$ scattering lengths~\cite{CuspEPJC}:
\bea
a_0^0-a_0^2 &=& 0.2571 \pm 0.0048_{\rm stat} \pm 0.0025_{\rm syst} \pm 0.0014_{\rm ext} ~, \nonumber\\
a_0^2 &=& -0.024 \pm 0.013_{\rm stat} \pm 0.009_{\rm syst} \pm 0.002_{\rm ext}  ~.\label{eq:results}
\eea
Comparing Eq.~\eqref{eq:results} to Eq.~\eqref{eq:results_sans}, we see that radiative corrections
decrease the extracted central value for $a_0^0-a_0^2$ by nearly 10\%, so their inclusion turns out to be
absolutely essential at this accuracy.

\begin{figure}
\includegraphics[width=\linewidth]{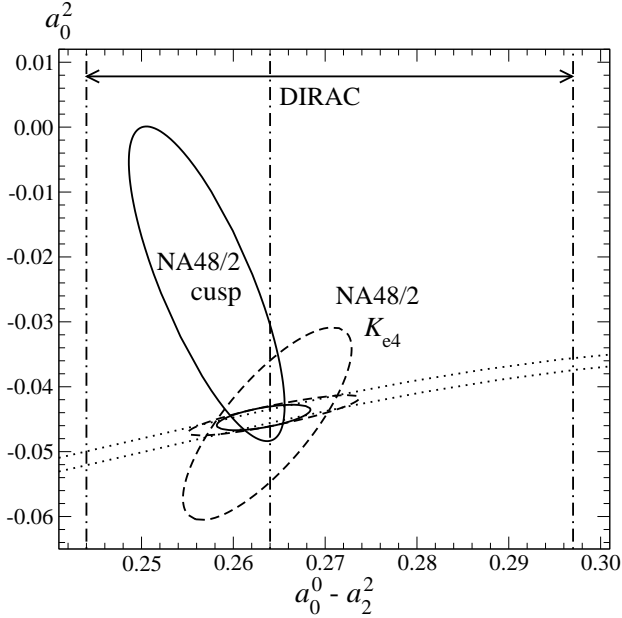}
\caption{Combined experimental results on the S-wave $\pi\pi$ scattering lengths
from the pionium width as obtained by the DIRAC collaboration~\cite{DIRAC}, $K_{e4}$~\cite{Batley:Ke4}, and the cusp
in $K^\pm\to\pi^0\pi^0\pi^\pm$~\cite{CuspEPJC}, the latter two as measured by the NA48/2 collaboration. 
For details, see text. Data obtained from Ref.~\cite{CuspEPJC}.}
\label{fig:a0a2ellipse}
\end{figure}
In Fig.~\ref{fig:a0a2ellipse}, we show the combined experimental results on the S-wave
$\pi\pi$ scattering lengths from modern precision determinations.
Shown are the ellipse (full) extracted from the cusp analysis in $K^\pm\to\pi^0\pi^0\pi^\pm$ 
described here~\cite{CuspEPJC},
the NA48/2 results from $K_{e4}$ decays~\cite{Batley:Ke4} (dashed) using theoretical input on 
isospin-breaking corrections~\cite{CGR:Ke4}, and the constraint from the pionium lifetime
obtained by the DIRAC experiment~\cite{DIRAC} as the vertical dash-dotted band.
The narrow dotted band shows the correlation between $a_0^0$ and $a_0^2$ dictated by the 
relation of $\bar\ell_4$ to the scalar radius of the pion~\cite{CGL2}, 
the smaller ellipses are fits to the two  NA48/2 experiments using the chiral perturbation theory constraint.
Altogether, a most impressive agreement between the different experiments as well as
experiment and theory has been achieved.

\section{On the accuracy of the extraction \\ \hspace{3.5mm}of \boldmath{$a_0^0-a_0^2$}}
\label{sec:accuracy}

An important ingredient yet missing to finally assess the accuracy of the extracted values 
for the $\pi\pi$ scattering lengths is a reliable estimate of the theoretical uncertainty
inherent in the representation of the amplitude.  In the first publication of an experimental 
cusp analysis~\cite{NA48},
a generic theoretical error of 5\% was assumed, following a suggestion made in Ref.~\cite{CI}, 
thus the theoretical input dominated the final uncertainty.
The following main points may be responsible for the theoretical error.
\begin{enumerate}
\item Radiative corrections.  These are now taken care of~\cite{radcorr}, we regard the remaining
uncertainty from higher-order radiative corrections as entirely negligible.
\item Isospin-breaking corrections in the matching relations.  These are small, the uncertainty
is estimated to be $\lesssim 1\%$, see Eq.~\eqref{eq:pipiMatch}. 
\item The effects of higher-order derivative interactions in the two-loop contributions,
i.e.\ terms of $\Order(a^2\eps^4)$ and higher.  Their impact is still under investigation~\cite{forthcoming},
although there are indications that the scattering lengths are very stable under such modifications
of the amplitude.
\item Higher loop contributions, starting at three loops $\Order(a^3\eps^3)$.
\end{enumerate}
For this last point, we wish to discuss the so-called \emph{threshold theorem}~\cite{fonda,Cabibbo,CGKR,radcorr}
(valid in the absence of photons).
It states that the coefficient of the leading cusp behavior (or $v_\pm(s_3)$) is proportional to the product
\beq 
T(K^+\to\pi^+\pi^+\pi^- )\big|_{\rm thr} \times T(\pi^+\pi^- \to \pi^0\pi^0)\big|_{\rm thr} ~, 
\eeq
where the second factor is just the combination of scattering lengths given in Eq.~\eqref{eq:pipiMatch},
and the ``threshold'' at which the first factor is to be evaluated refers to 
$s_1=4\Mc^2$, $s_2=s_3=(M_K^2-\Mc^2)/2$.  In other words, knowing the decay amplitude for the 
charged final state $T(K^+\to\pi^+\pi^+\pi^- )$ 
to $\Order(a^n)$ allows one to determine the dominant cusp strength of  $T(K^+\to\pi^0\pi^0\pi^+)$
at $\Order(a^{n+1})$.  As we have the full two-loop representation available for all $K\to 3\pi$ 
channels, we can estimate the size of the cusp at three loops by the expansion
\beq
 T(K^+ \to \pi^+ \pi^+ \pi^-) \bigl|_{\rm thr}
 \propto -1.0_{\rm tree} - 0.13\,i_{\rm 1-loop} + 0.014_{\rm 2-loop} ~,\label{eq:Nthresh}
\eeq
which suggests that the three-loop cusp will modify the leading (one-loop) effect by about 1.5\%.
This estimate is no substitute for a complete three-loop calculation, as it does not yield a
representation of $\Order(a^3)$ elsewhere in the decay region except near the cusp,
and neither does it contain information about subleading non-analytic behavior near threshold
(e.g.\ $\propto v_\pm^3(s_3)$).  Still, we regard Eq.~\eqref{eq:Nthresh} as a good indication
for the rate of convergence in the $K^+\to\pi^0\pi^0\pi^+$ amplitude.
\end{sloppypar}

\section{Cusps in other decays}

\subsection{\boldmath{$K_L\to3\pi^0$}, \boldmath{$\eta\to3\pi^0$}}\label{sec:KLeta}

\begin{sloppypar}
The mechanism generating the cusp in the $\pi^0\pi^0$ invariant mass distribution is rather generic
and only due to the final-state interactions between the pions.  We may therefore anticipate 
that other decays into two neutral pions plus a third particle will show a very similar cusp effect, 
the most obvious examples being $K_L \to 3\pi^0$ and $\eta\to3\pi^0$. 
The effect of the cusp in these channels has been investigated theoretically~\cite{CI,GPS,KL,Ditsche,Gullstrom},
and first efforts to see it experimentally have been reported both for $K_L\to 3\pi^0$~\cite{KTeV}
and $\eta\to 3\pi^0$~\cite{Adolph,Unverzagt,Prakhov} decays.
The main difference between these and $K^+\to\pi^0\pi^0\pi^+$, however, is the following.
\begin{figure}
\includegraphics[width=\linewidth]{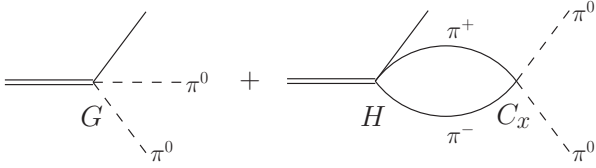}
\caption{Mechanism for cusp effects in generic decays into three final-state hadrons,
including two $\pi^0$.}
\label{fig:othercusps}
\end{figure}
As indicated in Fig.~\ref{fig:othercusps}, the extent to which the decay spectrum with respect to 
the invariant mass squared of the $\pi^0\pi^0$ pair is perturbed by the cusp effect does not only
depend on the charge-exchange scattering length as encoded in the coupling constant $C_x$, 
but strictly speaking, it is rather proportional to
\beq 
\frac{H}{G} \times C_x ~,
\eeq 
where $G$ and $H$ generically denote the coupling strengths to the ``neutral'' and ``charged'' 
final state, i.e.\ $\pi^0\pi^0$ and $\pi^+\pi^-$ plus a third meson, respectively.
In other words, the strength of the cusp depends crucially on the relative branching fractions
into the charged and neutral final states: the more the decay into charged pions is preferred,
the better the magnification of the effect in the spectrum.
\end{sloppypar}

It turns out that the ratio $|H/G|$ is very different for the different decays with three-pion 
final states.  While for the $K^+$ decays considered so far, it is approximately $2$, both 
for $K_L$ and $\eta$ it is closer to $1/3$, in other words the $K_L$ and $\eta$ prefer
to decay into $3\pi^0$.  
\begin{figure}\centering
\includegraphics[width=0.9\linewidth]{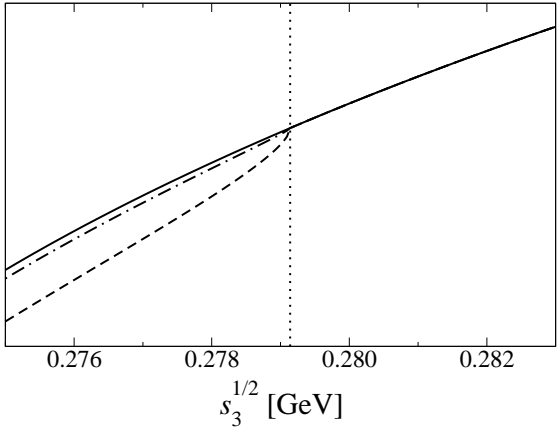} 
\caption{Sketches of the leading (one-loop) cusp effects on the decay spectra
for $K^+\to\pi^0\pi^0\pi^+$ (dashed line) and $K_L\to3\pi^0$ (dash-dotted line) 
in the vicinity of the $\pi^+\pi^-$ threshold, marked by the dotted vertical line.
The full line denotes the unperturbed spectrum without $\pi\pi$ rescattering.}
\label{fig:K+KLcusps}
\end{figure}
To illustrate the difference, we sketch the leading (one-loop, $\Order(a)$) cusps
for $K^+\to\pi^0\pi^0\pi^+$  and $K_L\to3\pi^0$ (the picture for $\eta\to 3\pi^0$ is very similar
to the latter case) in Fig.~\ref{fig:K+KLcusps}.  While the square-root-like structure
is clearly visible to the naked eye for the $K^+$ decay, it is much harder to discern in the 
case of the $K_L$.  For this reason, it is also much harder experimentally to achieve a precision
determination of $\pi\pi$ scattering lengths from an investigation of the cusps 
in $K_L\to 3\pi^0$ or $\eta\to3\pi^0$.  
More quantitatively, while the cusp effect reduces the number of events below the $\pi^+\pi^-$ threshold
in the $K^+\to\pi^0\pi^0\pi^+$ spectrum by about 13\%~\cite{MadigozhinCapri},
for e.g.\ $\eta\to3\pi^0$, the corresponding reduction amounts to only $1-2\%$~\cite{Gullstrom}.
So while probably no competitive scattering length determination from these channels seems
feasible in the near future, the cusp effect should at least be taken into account in
ongoing or future precision determinations of the $\eta\to3\pi^0$ Dalitz plot slope parameter 
$\alpha$~\cite{Adolph,Unverzagt,Prakhov}; compare also Ref.~\cite{Ditsche}.

\subsection{\boldmath{$\eta'\to\eta\pi^0\pi^0$}}

\begin{sloppypar}
At least from a theoretical perspective, much more promising in this respect is the decay
$\eta'\to\eta\pi^0\pi^0$~\cite{etaprime}.  With the $\eta'$ and the $\eta$ both being
particles of isospin 0, it is obvious that in the approximation of isospin conservation,
the $\pi\pi$ pair is produced with total isospin 0, which immediately shows that
the amplitude for $\eta'\to\eta\pi^+\pi^-$ is enhanced compared to the $\eta'\to\eta\pi^0\pi^0$
one by a factor of $-\sqrt{2}$ (the sign is according to the Condon--Shortley phase convention).
We therefore expect a cusp much more prominent than in $K_L,\,\eta\to3\pi^0$, if not quite as
pronounced as in $K^+\to\pi^0\pi^0\pi^+$.
From the experimental perspective, the upcoming high-statistics $\eta'$ experiments 
at ELSA~\cite{Beck}, MAMI-C~\cite{MAMI-C1,MAMI-C2,MAMI-C3}, WASA-at-COSY~\cite{WasaAdam,WASA}, 
KLOE-at-DA$\Phi$NE~\cite{KLOElett,KLOE}, or BES-III~\cite{BES} are expected
to increase the data basis on $\eta'$ decays by orders of magnitude, so an investigation
of the cusp effect in this channel seems very promising.

What makes this channel somewhat different from those investigated so far is the presence
of the $\eta$ in the final state, and hence of $\pi\eta$ rescattering as a new ingredient
to final-state interactions.
There is no experimental information on $\pi\eta$ threshold parameters, and it turns out
that chiral symmetry constrains these quantities only very badly~\cite{BKM:pieta,Kolesar}:
the $\Order(p^4)$ corrections to the current algebra prediction of the S-wave scattering length,
for example, can easily be as big as or bigger than the leading order.  
The one thing that chiral perturbation theory \emph{does} seem to predict reliably is the 
fact that $\pi\eta$ threshold parameters are systematically smaller than the $\pi\pi$ ones.
In conventions comparable to those chosen in $\pi\pi$ scattering (see Ref.~\cite{etaprime} 
for details), the S-wave $\pi\eta$ scattering length is given at leading order by
\beq
\bar a_0 = \frac{\Mpi}{96\pi F_\pi^2} + \Order(\M^4) ~,
\eeq
which compared to $a_0^0$, see Eq.~\eqref{eq:a00}, is smaller by a factor of 21.
We therefore expect the effect of the $\pi\eta$ final-state interactions to be significantly
smaller than that of $\pi\pi$ rescattering.
\end{sloppypar}

For an investigation of the impact of the $\pi\eta$ threshold parameters 
on the cusp effect in $\eta'\to\eta\pi^0\pi^0$, we vary them in a sensible range, 
suggested by various sets of next-to-leading order low-energy constants.
\begin{figure}
\includegraphics[width=\linewidth]{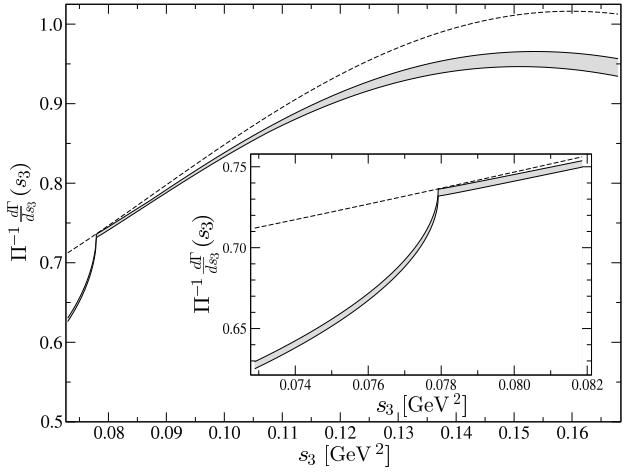}
\caption{Decay spectrum for $\eta'\to\eta\pi^0\pi^0$, divided by pure phase space.
The insert focuses on the cusp region around the $\pi^+\pi^-$ threshold.
The dashed line corresponds to the tree result, the gray band shows the full result to two loops, 
with the uncertainty due to the variation of the $\pi\eta$ threshold parameters.  
Figure taken from Ref.~\cite{etaprime}.}
\label{fig:etaprime}
\end{figure}
In Fig.~\ref{fig:etaprime}, we show the decay spectrum for $\eta'\to\eta\pi^0\pi^0$
with respect to the invariant mass of the $\pi^0\pi^0$ pair, comparing the spectrum
calculated from the tree-level amplitude to that given by the full two-loop result.
The tree-level couplings are fixed by (the central values of) the Dalitz plot parameters
determined in Ref.~\cite{VES}, and we assume isospin symmetry between these couplings 
for the $\eta'\to\eta\pi^0\pi^0$ and $\eta'\to\eta\pi^+\pi^-$ channels.
There is a very clear signal of the cusp effect below the $\pi^+\pi^-$ threshold,
plus a significant deviation between tree and two-loop spectrum mainly due to
$\pi\pi$ final state interactions at large $s_3=M_{\pi^0\pi^0}^2$.  
The width of the band gives an indication of the size of $\pi\eta$ rescattering effects,
which however hardly affect the cusp region.  

\begin{figure}
\includegraphics[width=\linewidth]{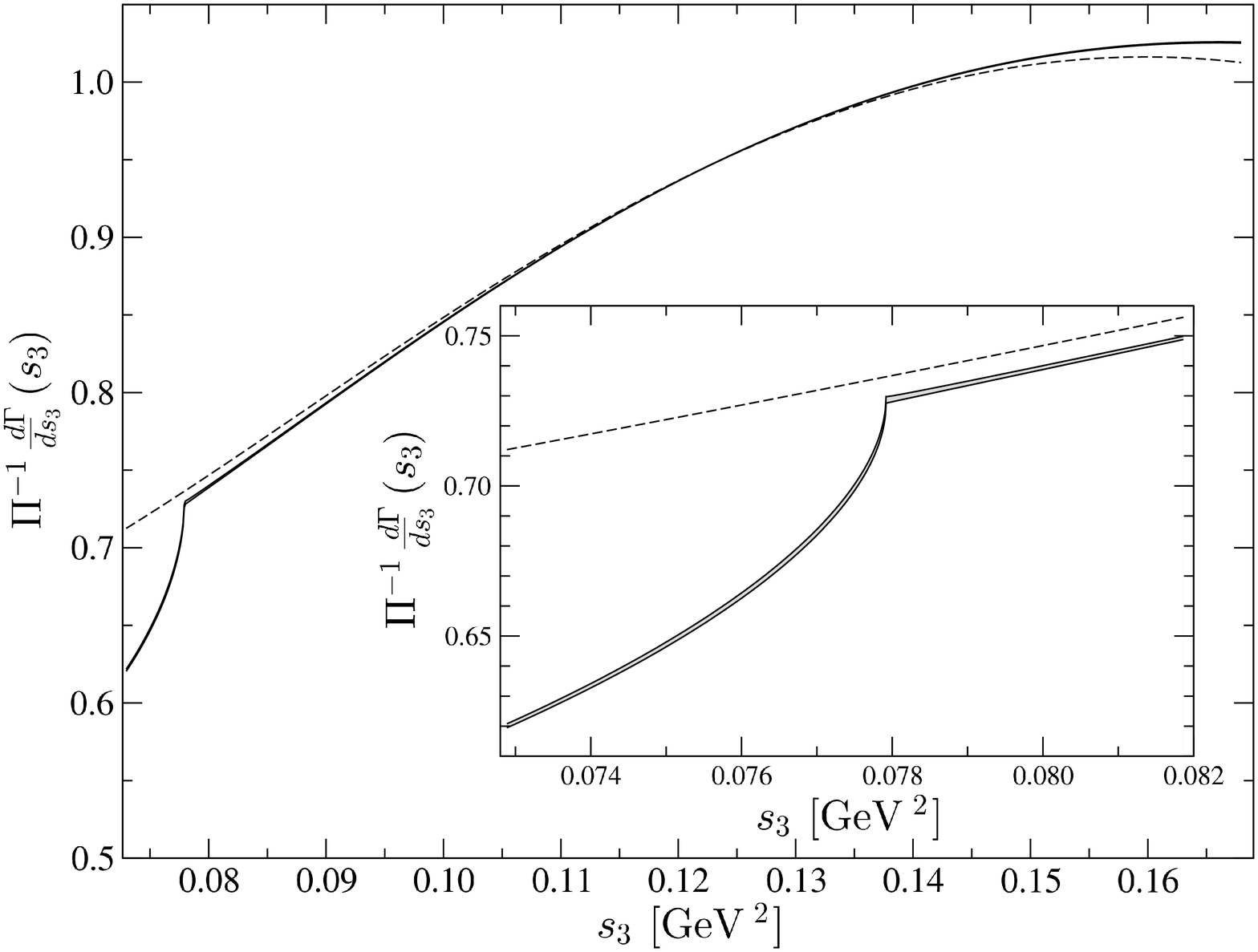}
\caption{Decay spectrum for $\eta'\to\eta\pi^0\pi^0$ as in Fig.~\ref{fig:etaprime}, 
after renormalization of tree couplings in order to  reproduce the Dalitz plot parameters
from Ref.~\cite{VES} with the full amplitude.  Figure taken from Ref.~\cite{etaprime}.}
\label{fig:etaprimeREN}
\end{figure}
Obviously, the Dalitz plot parameters of the distribution including loop corrections in
Fig.~\ref{fig:etaprime} are not identical to the input parameters any more.  In 
Fig.~\ref{fig:etaprimeREN}, we have therefore renormalized the tree-level couplings
in such a way that the \emph{full} amplitude reproduces the Dalitz plot expansion
as measured in Ref.~\cite{VES}.  The result is very striking, as the by far largest
part of the final-state interactions above the $\pi^+\pi^-$ threshold can be absorbed
into such a redefinition of the tree-level parameters.  In particular, hardly any
effect of the variation of $\pi\eta$ threshold parameters is visible any more.
On the other hand, the prediction of the cusp seems extremely stable.

A remarkable feature of Figs.~\ref{fig:etaprime}, \ref{fig:etaprimeREN} is the fact that
there seems to be hardly any trace left of what we discussed as the ``two-loop cusp''
in $K^+\to\pi^0\pi^0\pi^+$ decays, i.e.\ a square-root-like behavior \emph{above}
the $\pi^+\pi^-$ threshold.  This observation bears up against closer scrutiny:
numerically we find that the cusp above threshold in $\eta'\to\eta\pi^0\pi^0$ is 
suppressed by about a factor of 250 compared to the leading, $\Order(a)$ cusp.
The explanation for this suppression can be found with the help of the threshold
theorem again, see Sect.~\ref{sec:accuracy}, and it turns out to be the result
of residual approximate isospin symmetry between the amplitudes for 
$\eta'\to\eta\pi^0\pi^0$ and $\eta'\to\eta\pi^+\pi^-$~\cite{etaprime}.  So even if we allowed for
small isospin breaking in the tree-level couplings (which was neglected here), 
a strong relative suppression would persist.  

Finally, we can estimate the size
of a potential three-loop cusp in analogy to Sect.~\ref{sec:accuracy}, which turns
out not to be suppressed by similar arguments, although, naturally, by the high
power of scattering lengths involved.  In this case, we find that the cusp
of $\Order(a^3)$ should reduce the leading $\Order(a)$ cusp by about 0.5\%~\cite{etaprime}.
So in contrast to $K^+\to\pi^0\pi^0\pi^+$ decays, for the description of which
the $\Order(a^2)$ cusp turned out to be absolutely necessary, in the case of
$\eta'\to\eta\pi^0\pi^0$  the singularity is entirely dominated by the leading, 
one-loop rescattering term.

\subsection{The role of \boldmath{$\pi\eta$} interactions in \boldmath{$\eta'\to\eta\pi^0\pi^0$}}

\begin{figure*}
\includegraphics[width=\linewidth]{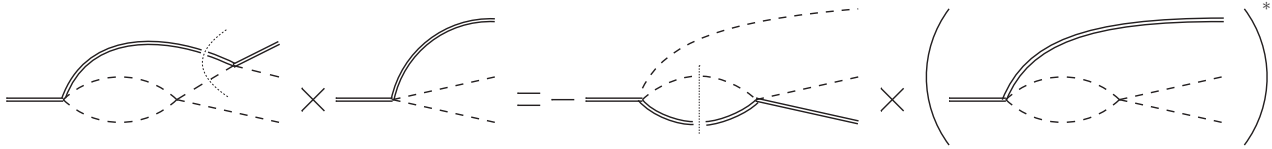}
\caption{Visualization of the cancellation of threshold cusps in $\eta'\to\eta\pi^0\pi^0$
at $s_1=(M_\eta+\Mn)^2$. Double lines denote both $\eta'$ and $\eta$ fields, dashed lines
stand for neutral pions.  Figure taken from Ref.~\cite{etaprimeCD09}.}
\label{fig:pietacanc}
\end{figure*}

The cusp effect in the $\pi^0\pi^0$ invariant mass spectrum of the decay
$\eta'\to\eta\pi^0\pi^0$ is dominantly (and, as we have seen above, practically completely)
due to $\pi\pi$ final-state rescattering.  An even more interesting question, however, may
be whether there is access to information on the $\pi\eta$ threshold parameters, too, 
in this decay.  The observation in the last section that at least a large part
of their effect can be absorbed in a redefinition of the tree-level couplings shows that
this is not a trivial endeavor, and at least not in the center of the Dalitz plot.

The obvious question to ask is whether there is an interesting non-analytic behavior
near the $\pi\eta$ thresholds, i.e.\ for $s_1$ / $s_2$ close to $(M_\eta+M_\pi)^2$.
In contrast to what makes the cusp at the $\pi^+\pi^-$ threshold so special, we
obviously cannot go \emph{below} threshold and really see a cusp as a change in slope
below vs.\ above a certain kinematical point; however, we still may look for 
square-root-like behavior at the border of the Dalitz plot, difficult as it would be
to investigate such a phenomenon experimentally.  
From what we have learnt so far, such a cusp above threshold would have to be 
a two-loop effect and therefore is expected to be small.  
However, things turn our to be even worse: with the use of the threshold theorem again,
applied now to the threshold $s_1=(M_\eta+M_\pi)^2$, one can show that the interference
of genuine two-loop graphs with the tree-level amplitude (both real) is always 
exactly cancelled by the product of two corresponding one-loop graphs (both imaginary
in our formalism), such that no square-root behavior survives
in the squared amplitude~\cite{etaprimeCD09}.  
This cancellation is shown schematically in Fig.~\ref{fig:pietacanc} for a specific
set of graphs, and it can be shown to persist for all diagrams up to two loops. 
We therefore have to conclude that, with the methods described here, we cannot identify
a method to extract $\pi\eta$ scattering lengths in a similar fashion 
as the cusp effect allows for the $\pi\pi$ ones.

\section{Summary and conclusions}

Non-relativistic effective field theory provides a systematic framework for an analysis
of the cusp phenomenon and pion--pion scattering lengths in $K^+\to\pi^0\pi^0\pi^+$ decays.
The representation of the decay amplitude is calculated in a combined expansion
in a non-relativistic parameter $\eps$ and $\pi\pi$ threshold parameters $a$, 
which is currently available up to $\Order(\eps^4,a\eps^5,a^2\eps^4)$.  
In order to match the enormous experimental accuracy achieved by the NA48/2
collaboration theoretically, radiative corrections have to be included.
The effect of the latter on the $\pi\pi$ scattering lengths is surprisingly large,
as photon effects modify the analytic structure of the decay amplitude near
the $\pi^+\pi^-$ threshold.
Similar cusp phenomena are also present in other decays such as $K_L\to3\pi^0$
or $\eta\to3\pi^0$, where they are however far less prominent and much harder 
to use for a precision determination of scattering lengths.
More promising in that respect is the decay $\eta'\to\eta\pi^0\pi^0$, 
which, on the other hand, seems not to offer easy access to $\pi\eta$
threshold parameters.

\section*{Acknowledgements}

\begin{sloppypar}
I am very grateful to my coworkers Moritz Bissegger, Gilberto Colangelo, Andreas Fuhrer, J\"urg Gasser,
Akaki Rusetsky, and Sebastian Schneider for the most fruitful collaborations 
leading to the results presented here,
and to J\"urg Gasser for useful comments on this manuscript.
I furthermore wish to thank Luigi Di Lella and Dmitri Madigozhin for providing
me with the data for Figs.~\ref{fig:expcusp} and \ref{fig:a0a2ellipse}.
We acknowledge the support of the European Community Research Infrastructure
Integrating Activity ``Study of Strongly Interacting Matter'' 
(acronym HadronPhysics2, grant agreement No.~227431) under the Seventh Framework Programme of the EU.
Work supported in part by DFG (SFB/TR 16, ``Subnuclear Structure of Matter'') and
by the Helmholtz Association through funds provided to the virtual 
institute ``Spin and strong QCD'' (VH-VI-231).
\end{sloppypar}

\end{document}